\documentclass[fleqn,usenatbib]{mnras}

\usepackage{newtxtext,newtxmath}
\usepackage[T1]{fontenc}
\usepackage{ae,aecompl}

\usepackage{graphicx}	% Including figure files
\usepackage{amsmath}	% Advanced maths commands
\usepackage{amssymb}	% Extra maths symbols

\title[Thawing the frozen-in approximation in TDEs]{Thawing the frozen-in approximation: implications for self-gravity in deeply plunging tidal disruption events}

\author[Steinberg, Coughlin, Stone, Metzger]{
Elad Steinberg\thanks{E-mail: es3640@columbia.edu},
Eric R.~Coughlin\thanks{Einstein Fellow},
Nicholas C.~Stone,
Brian D.~Metzger
\\
Columbia Astrophysics Laboratory and Department of Physics, Columbia University, New York, NY 10027, USA\\
}

\date{Accepted XXX. Received YYY; in original form ZZZ}

\pubyear{2019}
\begin{document}
\label{firstpage}
\pagerange{\pageref{firstpage}--\pageref{lastpage}}
\maketitle

% Abstract of the paper
\begin{abstract}
The tidal destruction of a star by a massive black hole, known as a tidal disruption event (TDE), is commonly modeled using the "frozen-in" approximation. Under this approximation, the star maintains exact hydrostatic balance prior to entering the tidal sphere (radius $r_{\rm t}$), after which point its internal pressure and self-gravity become instantaneously negligible and the debris undergoes ballistic free fall.  We present a suite of hydrodynamical simulations of TDEs with high penetration factors $\beta \equiv r_{\rm t}/r_{\rm p} = 5-7$, where $r_{\rm p}$ is the pericenter of the stellar center of mass, calculated using a Voronoi-based moving-mesh technique.   
We show that basic assumptions of the frozen-in model, such as the neglect of self-gravity inside $r_{\rm t}$, are violated.  Indeed, roughly equal fractions of the final energy spread accumulate exiting and entering the tidal sphere, though the frozen-in prediction is correct at the order-of-magnitude level.  We also show that an $\mathcal{O}(1)$ fraction of the debris mass remains transversely confined by self-gravity even for large $\beta$ which has implications for the radio emission from the unbound debris and, potentially, for the circularization efficiency of the bound streams.    

\end{abstract}

% Select between one and six entries from the list of approved keywords.
% Don't make up new ones.

%%%%%%%%%%%%%%%%%%%%%%%%%%%%%%%%%%%%%%%%%%%%%%%%%%

%%%%%%%%%%%%%%%%% BODY OF PAPER %%%%%%%%%%%%%%%%%%

\section{Introduction}
 Tidal disruption events (TDEs) occur when a star of mass $M_\star$ and radius $R_\star$ comes within a distance $r_{\rm t} \simeq R_\star\left({M_{\bullet}}/{M_\star}\right)^{1/3}$ of a supermassive black hole (SMBH), where $M_{\bullet}$ is the black hole mass. Within this distance, the tidal field of the black hole -- which stretches the star radially and compresses it perpendicularly -- surmounts the self-gravity of the star, which results in its destruction (e.g., \citealt{hills75, rees88}). This ability of a black hole to tidally destroy a star was validated with early simulations \citep{bicknell83, evans89}, and more recent numerical work has investigated the dependence of the disruption physics on the stellar composition and the orbital parameters \citep{lodato09, guillochon09, guillochon13, mainetti17, lawsmith17,movingmesh}. 

In addition to numerical simulations, TDEs can be qualitatively (and, to a lesser extent, quantitatively) well-understood through the analytic, ``frozen-in'' approximation (\citealt{lacy82,rees88}). This approximation assumes that the disrupted star maintains perfect hydrostatic balance prior to reaching the tidal radius and is thereafter ``destroyed,'' meaning that the pressure and self-gravity of the star are negligible past this point, and the fluid parcels comprising the star therefore trace out ballistic orbits in the black hole potential (or the relativistic analog; e.g., \citealt{Kesden12}). Using this formalism, one can derive the tidal spread in the energy imparted by the black hole, the return timescale of the most tightly bound debris, and the fallback rate (i.e., the rate at which matter returns to pericenter) as a function of the impact parameter \citep{stone13}, stellar structure \citep{lodato09}, and stellar spin \citep{stone13,golightly19}.

While the frozen-in approximation is useful for obtaining a rough understanding of TDEs, it ignores a number of physical effects that can be important for modifying the tidal disruption picture. For one, the disrupted stellar debris quickly recedes beyond the tidal radius -- where the tidal field and the star's original self-gravity are equal -- and the time taken to return to pericenter is on the order of weeks to months for Solar-like stars and black hole masses $M_{\bullet} \sim 10^{6} M_{\odot}$. Ignoring the self-gravity of the tidally-disrupted debris is therefore not an obviously reasonable assumption, and simple analytic arguments suggest that self-gravity may reconfine the debris in directions transverse to its motion \citep{kochanek94}. In support of this notion, simulations have shown that bound cores can recollapse out of the disrupted material \citep{guillochon13}, and the tidally-disrupted stream itself can, depending on the equation of state of the gas, be gravitationally unstable and condense into small-scale knots \citep{coughlin15,coughlin16a,coughlin16b}.

As another example, the compression of the gas near pericenter for large $\beta = r_{\rm t}/r_{\rm p}$, where $r_{\rm p}$ is the pericenter distance of the center of mass, becomes extreme. Specifically, for an ideal gas with a polytropic index of $\gamma=5/3$, the compression ratio $\rho_{\rm max}/\rho_{\star}$, where $\rho_{\rm max}$ and $\rho_{\star}$ are the maximum density and original stellar density, respectively, is predicted to scale as %the very steep function of $\beta$
$\rho_{\rm max}/\rho_{\star} \sim \beta^3$ \citep{carter83}. In these high-$\beta$ encounters, it is plausible that the increase in the pressure or self-gravity in response to the extreme tidal squeezing could modify the predictions of the frozen-in approximation. Indeed, while simulations and analytic estimates suggest that the frozen-in approximation is roughly valid in this regime of modest to large $\beta$ \citep{guillochon13,stone13}, a rigorous assessment of its accuracy has not been performed.

In this Letter, we perform a numerical study of the impact of large $\beta$ on the predictions of the frozen-in approximation. %We employ the Voronoi based moving mesh technique, which allows to accurately capture the fluid dynamics as well as allows large compression ratios to be resolved. 
We use the 3D version of the open-source code \textsc{RICH} \citep{RICH}, which solves the compressible Euler equations on a moving Voronoi mesh with a finite volume Godunov scheme. The self-gravity of the star is calculated using a tree method including up to quadrupole moments. All the runs are performed in the center of mass frame of the star and the tidal field's Newtonian acceleration is added to the calculation (we ignore special and general relativistic effects).  We use an ideal gas equation of state with an adiabatic index $\gamma=5/3$.  For the initial stellar structure we adopt a Lane-Emden polytrope ($P \propto \rho^{\Gamma}$) with $\Gamma \equiv 1 + 1/n$ for two cases corresponding to a convective ($n = 1.5$) and radiative ($n = 3$) star.  The mass of the SMBH is taken to be $M_{\bullet} = 10^6M_\odot$ and the star has a Solar radius ($R_{\star} = R_{\odot}$) and mass ($M_\star = M_{\odot}$). % equal to those of the Sun.  
We perform simulations for orbital trajectories corresponding to relatively large penetration factors of $\beta = r_{\rm t}/r_{\rm p} = 5-7$, where $r_{\rm t}$ ($r_{\rm p}$) is the tidal (pericenter) radius.  We adopt a resolution of roughly $2\cdot 10^7$ cells and all of the runs start at an initial distance of $r=3\,r_\textrm{t}$.

%Give background
\section{Results}

\begin{figure}
  \includegraphics[width=\linewidth]{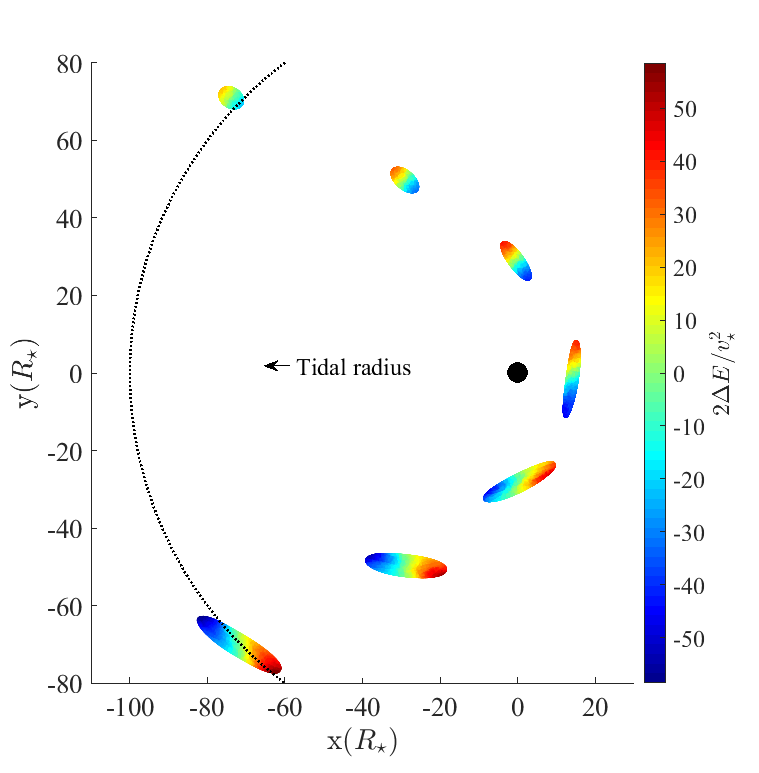}
  \caption{Time evolution of the star as it passes through the tidal sphere for a run with $\beta=7$ and $n=1.5$.  We show the gas above a density threshold that encompasses $50\%$ of the total stellar mass.  Color denotes the specific orbital energy of the gas, normalized to the square of the stellar escape velocity. The size of the star has been increased by a factor of 5 in order to show detail.}
  \label{fig:lab}
\end{figure}

Figure \ref{fig:lab} shows an example of one of the deeply-penetrating orbits in the lab frame, illustrating the evolution of the fluid within the orbital plane of the incoming star.  As the star enters the tidal sphere, it becomes increasingly distorted by the tidal field of the SMBH, being stretched in the radial direction and compressed orthogonally. Following the pericenter passage, the disrupted star continues to expand preferentially in the radial direction, which results in the formation of a tidally-disrupted ``stream'' of stellar debris. 

%\begin{figure}
%  \includegraphics[width=\linewidth]{energy_full}
%  \includegraphics[width=\linewidth]{self_bound_scatter.png}
%  \caption{{\bf Top:} Blah Blah {\bf Bottom:} Orbital plane cross section showing the gas density at a snapshot in time as the stellar debris exits the tidal sphere for the $n=1.5$ stellar profile and orbit $\beta=7$.  The scale is in logarithm of gas density $(\rho/$g cm$^{-3})$ and distance are in units of the initial stellar radius.  A white contour encapsulates gas that is self-bound according to the criterion $\rho > \rho_{\bullet}$.}
%  \label{fig:lab}
%\end{figure}

%\begin{figure}
 % \includegraphics[width=\linewidth]{self_bound_scatter.png}
  %\caption{Orbital plane cross section showing the gas density at a snapshot in time as the stellar debris exits the tidal sphere for the $n=1.5$ stellar profile and orbit $\beta=7$.  The scale is in logarithm of gas density $(\rho/$g cm$^{-3})$ and distance are in units of the initial stellar radius.  A white contour encapsulates gas that is self-bound according to the criterion $\rho > \rho_{\bullet}$. \textbf{ERC: label for the color bar? and at what time is this? (nvm, I see that you said it above)}}
 % \label{fig:selfbound_scatter}
%\end{figure}

\begin{figure}
  \includegraphics[width=\linewidth]{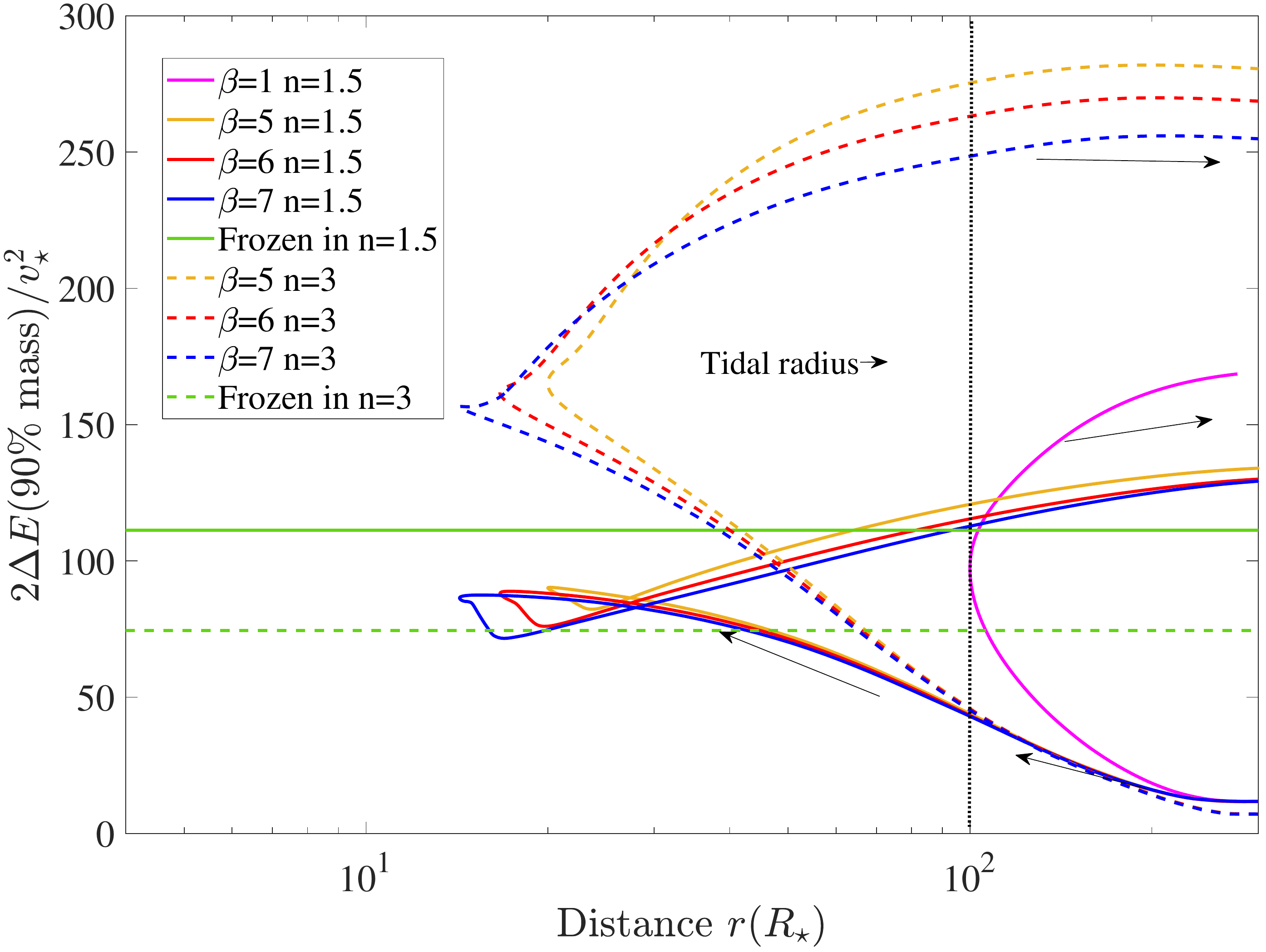}
  \caption{Evolution of the energy spread of the stellar material as a function of distance from the SMBH.  Solid(dashed) lines show cases corresponding to a $n=1.5(3)$ polytrope stellar structure, while different values of $\beta$ are shown as different colors.  For comparison, the energy spread predicted by the frozen-in approximation (Equation \ref{eq:fi}) for each polytrope is shown by a horizontal green line. Arrows show the direction of the star's center of mass.}
  \label{fig:de}
\end{figure}

Figure \ref{fig:de} shows the spread in the specific binding energy of the debris (i.e., kinetic plus potential energy) between the 5th and 95th percentiles by mass enclosed from the center of mass of the stream, as a function of the distance of the center of mass from the SMBH. The green curves give the predictions from the frozen-in approximation, being roughly 

\begin{equation}
    \Delta E \simeq \frac{\partial \Phi}{\partial r}\Delta R = \frac{2GM_{\bullet}}{r_{\rm t}^2}R_\star = \left(\frac{M_\bullet}{M_\star}\right)^{1/3}v_\star^2,
\label{eq:fi}
\end{equation}
where $\Phi = -GM_\bullet/r$ is the potential of the black hole and $v_\star = \sqrt{2GM_\star/R_\star}$ is the escape speed from the stellar surface; the slight differences between the green curves in Figure \ref{fig:de} arise from the structure of the star, which we incorporated following the approach of \citet{lodato09}. The other curves show the results for different stellar profiles and pericenters as indicated in the legend, while the arrows represent the direction of motion of the center of mass of the star (i.e., arrows pointing to the left indicate that the distance to the black hole is decreasing and vice versa). 

We see from this figure that, instead of displaying a step-like change as the star crosses the tidal radius (as is assumed by the frozen-in model), the energy spread evolves continuously while the star is within -- and slightly outside of -- the tidal sphere of the black hole. Specifically, as each star plunges toward the SMBH, its energy spread increases to approximately half of its peak value by the time the center of mass reaches pericenter, and the remaining half is nearly achieved by the time the center of mass recedes back to the tidal radius. Moreover, while the final energy spread is comparable to the frozen-in prediction for the $n = 1.5$ polytropes, and is nearly independent of $\beta$ ($\beta$-independence being another prediction of the frozen-in approximation), the $n = 3$ polytropes maintain an energy spread that is larger than the frozen-in value by a factor of $\sim 5$. There is also a more noticeable dependence on $\beta$ for these encounters. 

\begin{figure}
  \includegraphics[width=\linewidth]{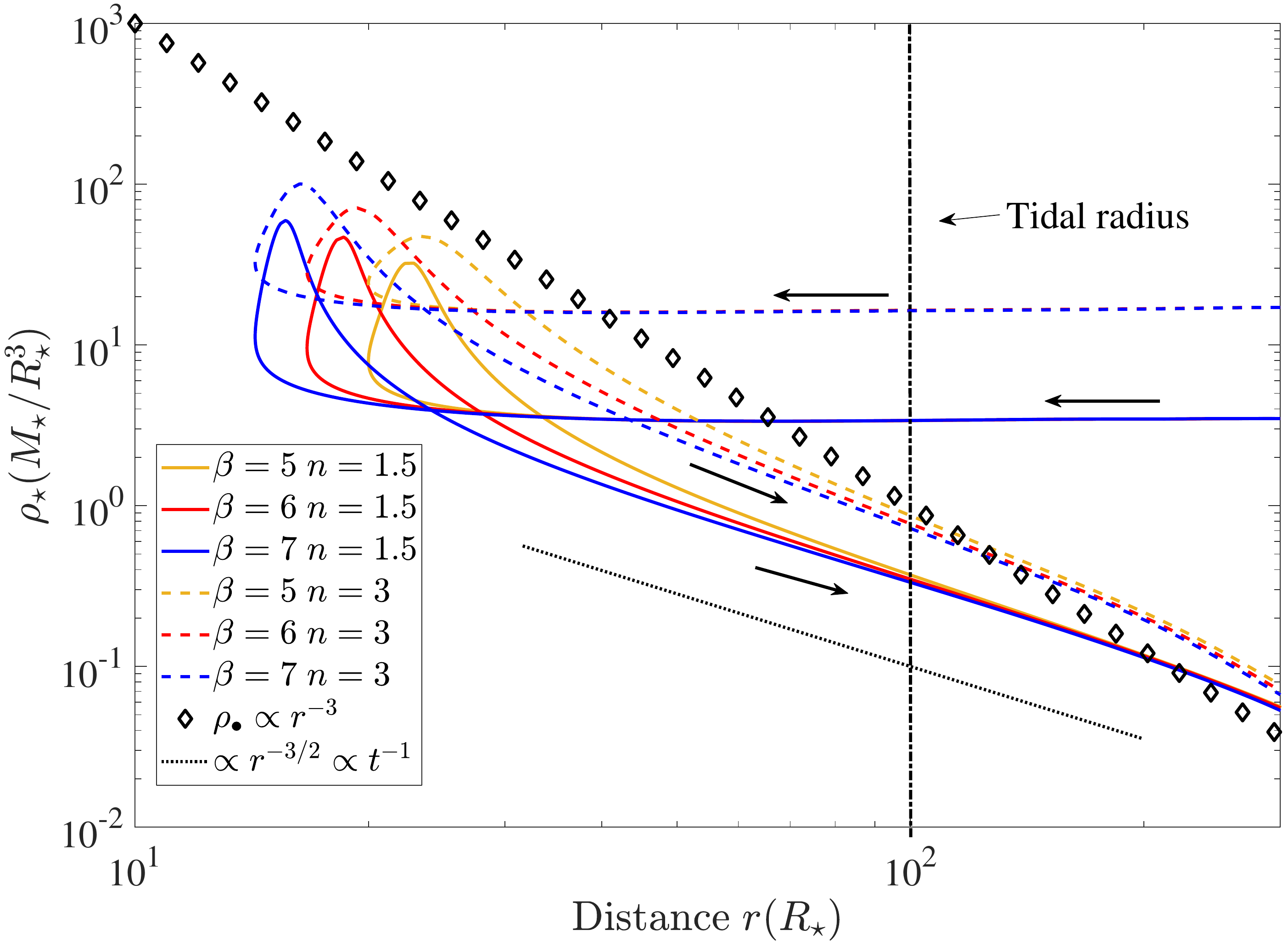}
  \includegraphics[width=\linewidth]{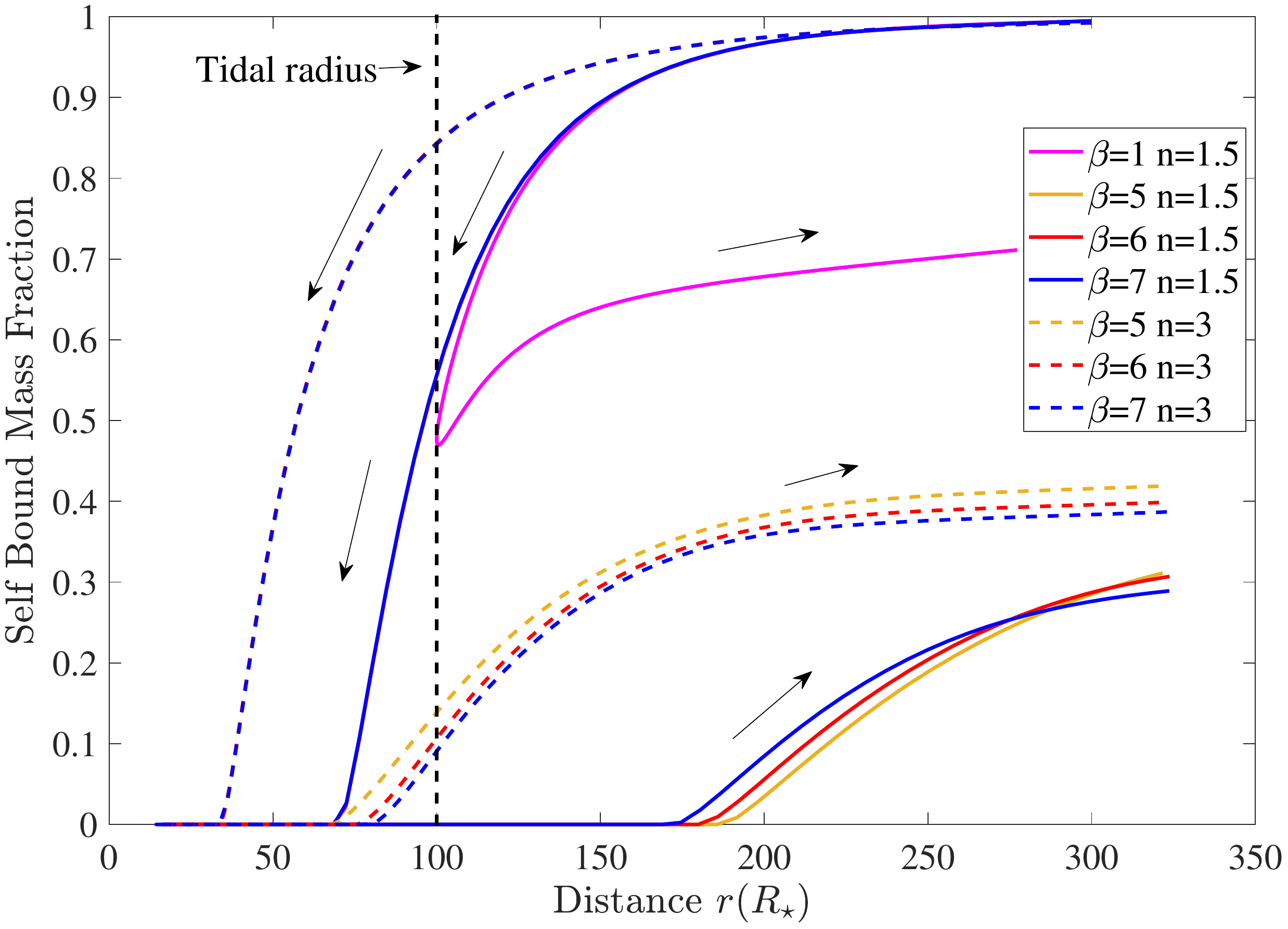}
  \caption{{\bf Top:} Stellar density, $\rho_{\star}$, interior to the 50\% mass percentile (colored lines) compared to the SMBH ``density'' $\rho_{\bullet} \equiv M_{\bullet}/(2\pi r^3)$ (open diamonds), as a function of distance $r$ from the SMBH to the center of mass of the star.  Solid lines show the evolution of a $n=1.5$ polytrope and the dashed lines for an $n=3$ polytrope. Arrows show the direction of temporal evolution. {\bf Bottom:} Fraction of the stellar mass that obeys $\rho > \rho_{\bullet}$, which is therefore bound in the vertical direction due to self-gravity, as a function of $r$.}
  \label{fig:selfbound}
\end{figure}

The bottom panel of Figure \ref{fig:selfbound} illustrates the fraction of the stellar debris that remains self-gravitating, which, following \citet{kochanek94} and \citet{coughlin16b}, we define according to the criterion $\rho \ge \rho_{\bullet} \equiv M_{\bullet}/(2\pi r^3)$, where $\rho$ is the density of the disrupted debris; if the density at any point in the stream satisfies this inequality, then the self-gravity of the gas exceeds the tidal field of the black hole (we note, however, that this criterion does not incorporate the motion of the gas in the transverse direction; we return to this point in Section \ref{sec:self-gravity}). This figure demonstrates that, as the stars plunge within the tidal sphere, the importance of self-gravity is reduced and the self-gravitating fraction approaches zero. This result is in agreement with the expectation that no part of the stream should be self-gravitating if the pericenter of the encounter satisfies $\beta \gtrsim \left(\rho_{\rm c}/\rho_\star\right)^{1/3}$, where $\rho_{\rm c}$ ($\rho_{\star}$) is the central (average) stellar density, as in this case the density of the black hole overwhelms even the densest region of the star \citep{kochanek94}. 

Surprisingly, however, once the center of mass of the star passes through pericenter and recedes out to a modest fraction of the tidal radius, the self-gravitating fraction begins to increase. Furthermore, the percentage of the gas that is self-gravitating is almost \emph{completely independent of $\beta$} for all $\beta \ge 5$, and there is only a slight dependence on the stellar structure. We also see that while the $n = 3$ simulations have levelled off in the amount of self-gravitating material contained within the stream by the time the simulations were terminated, the $n = 1.5$ simulations have not yet converged.  This finding suggests that self-gravity continues to be important at times long after the stream recedes beyond the tidal sphere \emph{even for these high-$\beta$ disruptions}.

\begin{figure}
 \includegraphics[width=\linewidth]{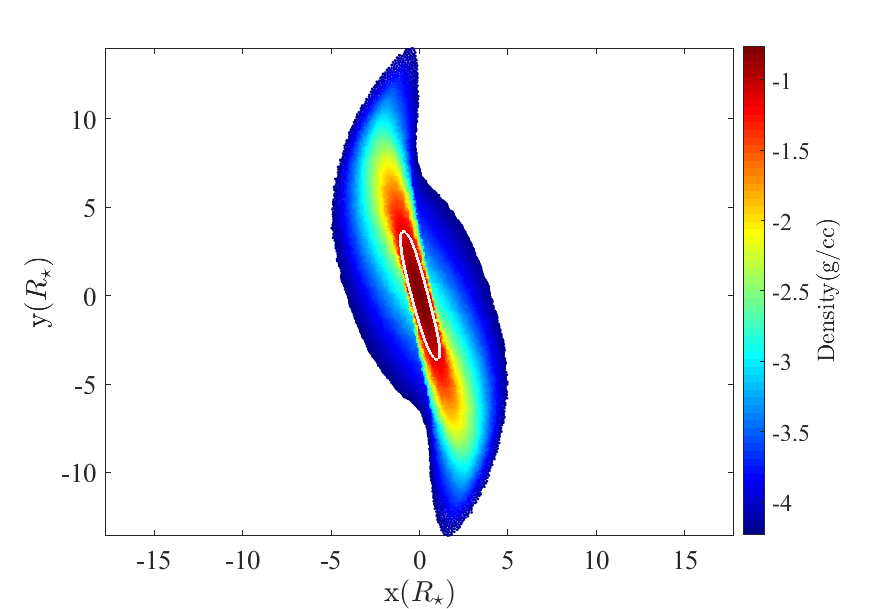}
 \caption{Orbital plane cross section showing the gas density at a snapshot in time as the stellar debris exits the tidal sphere for the $n=1.5$ stellar profile and orbit $\beta=7$.  The scale is in logarithm of gas density $(\rho/$g cm$^{-3})$ and distance are in units of the initial stellar radius.  A white contour encapsulates gas that is self-bound according to the criterion $\rho > \rho_{\bullet}$.}
  \label{fig:selfbound_scatter}
\end{figure}

Figure \ref{fig:selfbound_scatter} shows the stellar density within the orbital plane from the simulation with $n = 1.5$ and $\beta = 7$ at the time when the center of mass exits the tidal sphere, with colors indicating the logarithm of the density (in g cm$^{-3}$) as shown in the color bar. The white contour indicates the fraction of the stream within which the self-gravitating criterion is satisfied. This figure therefore shows that, while the outer extremities of the stream are dominated by the tidal shear of the black hole, self-gravity remains important for the central, cylindrical portion of the material. At late times, it thus follows that the stream possesses a ``spine'' of denser, self-gravitating material that is surrounded by a ``sheath'' of non-self-gravitating gas. 

\section{Discussion}
Here we provide a physical basis for the violation of the frozen-in nature of the energy spread and the self-gravitating nature of the stream.

\subsection{Energy Spread}

Approximate free-fall solutions show that as the star plunges through the tidal radius, it takes on a cylindrical shape, where the length of the star along the orbit, $L(t)$, and the cylindrical radius, $R(t)$, evolve according to $L(t)\propto 1/R(t)$ (e.g.~\citealt{stone13})\footnote{More precisely, the free-fall solutions approximate a prolate ellipsoid that follows $L(t) \propto 1/R(t)$ until it begins to exit the tidal sphere.}.  For simplicity, we assume the star's density along the cylinder remains uniform during its tidal evolution and that the stellar center of mass orbit can be approximated as purely radial (which is valid for high-$\beta$ encounters). The magnitude of the acceleration experienced by fluid elements at the ends of the cylinder due to self-gravity is then 

\begin{equation}
    a_{\rm sg} \approx \frac{2GM_\star}{R_\star^2}\left(1+\mathcal{O}\left(\frac{R_\star}{L(t)}\right)^2\right). \label{eq:asg}
\end{equation}
Since the stellar gas remains approximately adiabatic through pericenter passage, the acceleration due to the pressure gradient along the orbital motion is given by
\begin{equation}
    a_{P} \approx \frac{P_{\star}}{\rho_{\star}R_\star}\left(\frac{R_\star}{L(t)}\right)^{2-\gamma}\approx\frac{GM_\star}{R_\star^2}\left(\frac{R_\star}{L(t)}\right)^{2-\gamma}
\end{equation}
Since the length of the star along the orbit increases as $L(t)\propto 1/\sqrt{r}$ \citep{Sari2010,stone13}, where $r$ is the distance from the stellar center of mass to the SMBH, we find that 
\begin{equation}
    \frac{a_{\rm sg}}{a_{P}} \propto \begin{cases}
     r^{-1/6}& \gamma=5/3\\
     r^{-1/3}& \gamma=4/3
    \end{cases},
\end{equation}
i.e., the acceleration due to self-gravity %comes to 
dominates over that due to the pressure gradient.  %The power exerted by self-gravity on the star at radius $r < r_t$ is given by
%\begin{equation}
 %   P=a_{\rm sg}v_{\rm orb}=a_{\rm sg}\sqrt{\frac{2GM_{\bullet}}{r}} \label{eq:Psg}
%\end{equation}
%where $v_{\rm orb}$ is the orbital velocity. The resulting contribution to the energy spread of the debris at each decade in radius is then given by
The work done by self-gravity is then the integral of Equation \eqref{eq:asg} along the orbit, which generates a change in the specific energy of

\begin{equation}
    \Delta E \approx a_{\rm sg}r = \frac{2GM_\star}{R_\star^2}r\propto r. \label{eq:Esg}
\end{equation}
The fact that the total work done on the star is dominated by large $r$ implies that the majority of the energy spread occurs as the star travels from $r \approx r_{t}$ to roughly half that distance.

Figure \ref{fig:de} shows that, as each star plunges toward the SMBH, its energy spread increases continuously until pericenter and the majority of the increase occurs as the star reaches a distance of half the tidal radius, in agreement with the analytic considerations above.  The energy spread is noticeably higher for the cases of a more centrally concentrated star ($n = 3$ Lane-Emden profile) than the more homogeneous case of $n=1.5$. This arises from the fact that the self gravity is higher in regions of the star where most of the mass is located.  On the other hand, as mentioned earlier, the energy spread at the tidal radius is significantly less than that predicted by the frozen-in approximation (horizontal green lines).  This discrepancy arises because, even far away from the SMBH, the pressure gradient roughly balances self gravity.  The fact that tidal deformation begins even before the star enters the tidal radius allows the stellar material to acquire the kinetic energy needed to offset changes in its potential energy.

Figure \ref{fig:de} also shows that, as the star exits pericenter, its energy spread continues to grow. 
After pericenter passage, the star collapses in the vertical direction, a process which is halted once the pressure builds up to a sufficiently large value. The star then bounces back due to the pressure gradient, expanding rapidly in the vertical direction. Additionally, the star expands along its minor axis in the orbital plane, most noticeably once the star's height has become comparable with the length of the minor axis.  Since the minor axis of the star has a component parallel to the orbital motion \citep{stone13}, this expansion causes a further increase in the energy spread that manifests by the asymmetry in the energy distribution around the major axis (Fig.~\ref{fig:lab}).

Overall, the magnitude of the energy spread is comparable with the naive frozen-in approximation, $\Delta E\approx (M_{\bullet}/M_\star)^{1/3}v_\star^2$ (Equation \ref{eq:fi}). This result is in agreement with Equation \eqref{eq:Esg}, which shows (by setting $r = r_{\rm t}$) that the influence of self-gravity generates an energy spread that is on the order of the frozen-in prediction. %This is not a coincidence because the frozen-in approximation states that $\Delta E =2v_\star v_{\rm orb}(r=r_t)$, while we find that the spread is induced by acceleration due to the star's self-gravity acting over the dynamical timescale of the star (eq.~\ref{eq:Psg}).  This gives rise to an impulse of $\Delta v \approx a_{sg}t_\star\approx v_\star$.

For comparison, we also simulated a disruption with $\beta=1$ and $n = 3/2$. Perhaps surprisingly, the energy spread of the debris is actually larger for $\beta = 1$ than for the deeply penetrating cases with $\beta \gg 1$.  This finding can be understood by noting that %, for $\beta = 1$, %, self-gravity plays a larger role since, for a Keplerian parabolic orbit, 
the time required for the star to travel from $r=2r_t$ to $r=r_t$ -- and therefore the time over which self-gravity modifies the energy distribution, is smaller for larger $\beta$.

\subsection{Self-bound Fraction}
\label{sec:self-gravity}
%It is commonly assumed in the TDE literature that high $\beta$ events result in complete disruption of the star, leaving little of the gas self-gravitating following pericenter passage (Add citation). 
%Simple analytic arguments suggest that, if $\beta \gtrsim \left(\rho_{\rm c}/\rho_*\right)^{1/3}$, where $\rho_{\rm c}$ is the central density of the star and $\rho_*$ is its average density, self-gravity should not affect the evolution of the disrupted debris \citep{kochanek94}. In agreement with this argument, previous works have shown that for sufficiently high $\beta$ the star is completely destroyed and does not leave behind a bound core \citep{guillochon13, mainetti17}. However, even when there is no intact stellar remnant, the motion in the transverse direction (i.e. perpendicular to the orbital plane) can remain confined by the self gravity of the gas \citep{coughlin16a}. 
As noted above, a metric for measuring the importance of self-gravity in the transverse (i.e., non-radial) directions %compared to the tidal shear of the black hole 
is the ratio of the stream density $2\pi\rho$ to the SMBH density, $\rho / \rho_\bullet = 2\pi \rho r^3/M_{\bullet}$ \citep{kochanek94,coughlin16b}. For a $\gamma = 5/3$ gas and under the assumption that the dynamical acceleration of the stream width is small, if this ratio ever falls below unity at any segment of the stream, then that segment will never again become self-gravitating (although the energy from radiative recombination, which occurs weeks to months post-disruption, can further reduce the self-gravitating fraction; \citealt{kasen10,Guillochon+16}).%(although late-time injections of energy into the stream from e.g. recombination can later unbind it, as in \citealt{Guillochon+16}). 

Figure \ref{fig:selfbound} presents a clear violation of this prediction: for every TDE with $\beta > 1$, there exists a period where the \emph{entire stream} is dominated by the tidal shear of the black hole. Nonetheless, at approximately the time at which the stream exits the tidal sphere, the self-gravity of some fraction of the debris regains importance over the tidal field of the black hole. We suggest that the origin of this re-emergence of self-gravity is due to ``dynamical focusing'' within the orbital plane of the star, which is a consequence of the nearly-ballistic nature of the orbital motion of the gas parcels comprising the star as it passes through the tidal radius. Specifically, as pointed out by \citet{coughlin16a}, small differences in the acceleration of gas parcels within the plane result in the transverse compression of the stream.  In the absence of pressure, an in-plane caustic would occur where fluid elements cross. This transverse focusing then produces a higher density than would be predicted if the stream were in rough hydrostatic balance, which correspondingly allows the stream density to overtake that of the black hole at a later time.

In support of this argument, the top panel of Figure \ref{fig:selfbound} shows the density at the point within the stream where 50\% of the mass is contained\footnote{More specifically, integrating the density over the volume of each cell in the simulation from most to least dense, this is the point where one encloses 50\% of the total mass.} as a function of the Lagrangian distance of the center of mass. If the stream were in approximate hydrostatic balance, then the density would follow $\rho_\star \propto r^{-3}$ \citep{coughlin16b}, which is shown by the line of diamond symbols. When the center of mass of the stream recedes beyond the tidal radius, we see from this figure that the stream density resulting from disruptions of $n = 3$ polytropes and, to a lesser extent, $n = 1.5$ polytropes declines roughly proportionally to this scaling. However, at earlier times the stream density falls off significantly less steeply with radius than would be predicted by approximate hydrostatic balance and, as shown by the dotted line in this figure, is better matched by $\rho_\star \propto r^{-3/2}$. This finding suggests that there is some additional, dynamical evolution of the stream cross section that revitalizes the influence of self-gravity. %Owing to the fact that the stream expands radially as $\propto r^{2}$ at early times \citep{coughlin16b}, this scaling of the density implies that the stream cross-sectional area actually \emph{decreases} as $\propto r^{-1/2}$. This result suggests that during this initial phase of post-pericenter evolution, the stream is indeed compressed in the transverse direction, and this dynamical confinement enhances the self-gravitating nature of the stream {\bf ES: This needs to be emended a bit. I actually find that from pericenter upto the tidal radius the length of the gas hardly changes, this is because the length has two solutions, one that you state and another that goes like $\propto 1/\sqrt{r}$}.  

\section{Conclusions}
We presented a suite of simulations of deeply-penetrating TDEs as well as a simple toy model to explain our main findings.  While qualitatively the energy spread of the debris is similar to that predicted by the ``frozen-in" approximation (Equation \ref{eq:fi}), its physical origin arises from the continuing work done by self-gravity, quite unlike the impulse assumption underpinning the former.  In each tidal disruption, a significant fraction of the gas remains self-gravitating in the vertical direction by the end of the simulation, forming a gravitationally-confined ``spine" of dense gas surrounded by a dilute ``sheath" of shear-dominated debris.

Our results have several observable implications.  Owing to general relativistic precession, the apsidal angle of the pericenter of the returning debris stream is advanced by an additional amount, which results in the self-intersection of the stream (e.g., \citealt{rees88}). This stream-stream collision is thought to be the main mechanism by which the tidally-disrupted material dissipates kinetic energy and forms an accretion flow, and this notion has been substantiated with hydrodynamical simulations \citep{hayasaki13, shiokawa15,bonnerot16,hayasaki16, sadowski16}. For high-$\beta$ encounters, the self-intersection radius becomes smaller and the collision becomes stronger owing to the correspondingly larger relative velocity.
%and the diminished influence of nodal precession (i.e., precession out of the plane caused by the black hole spin does not have enough time to generate a substantial deflection of the outgoing stream). 
Our results imply that, because a significant fraction of the stream remains self-gravitating \emph{even for these high-$\beta$ encounters}, the collision strength for these more relativistic encounters is further augmented by the increased density. Therefore, we expect disc formation to be expedited by the self-gravitating nature of the stream in high-$\beta$ TDEs, so long as orbits remain coplanar.  On the other hand, self-gravitating debris streams present a much narrower geometric cross-section than do those moving ballistically, making it easier for the Lense-Thirring effect to prevent self-intersection in cases where the SMBH spin is misaligned from the stellar angular momentum \citep{kochanek94, hayasaki16}

The unbound gas that is not self-gravitating continues on a ballistic trajectory as it speeds away from the SMBH. The interaction of the this gas with the dense material typically found in galactic centers creates a bow shock %at the head of the unbound gas
that can lead to synchrotron emission (e.g.~\citealt{Guillochon+16,radio2019}).  The smaller cross section of the stream (resulting from self-gravity) implies that  %if it is self-bound means 
it will take significantly longer to sweep up ISM and decelerate than if it were unconfined \citep{Guillochon+16}, which naively generates a correspondingly reduced radio luminosity.  Even though we find that the self-gravitating fraction is high in many cases, even a small fraction of unconfined mass may be sufficient to power an observable radio signal \citep{krolik2016}. The smaller surface area of the vertically confined stream could also significantly reduce the emission line signatures from the unbound debris being irradiated by the TDE accretion flow (e.g.~\citealt{Strubbe&Quataert11}).

\citet{guillochon09} and \cite{breakout} estimate the brief flare of electromagnetic radiation produced by the shock created from tidal compression as it emerges from the outer layers of the star.  Although we observe in our simulations a shock propagating in the vertical direction after collapse, its power is significantly less than that assumed in these works.  This result makes the detection of X-ray emission from shock break-out in main-sequence stars even more challenging than these previous estimates. %estimated by these authors since its luminosity will be lower.
However, for giant stars, while the break-out signal would be less luminous, the larger stellar radius and reduced shock power estimation places its spectral peak more into the optical band, enhancing its detectability. A more detailed analysis of the shock breakout signal predicted from our simulations will be presented in future work (Steinberg et al.~2019, in prep).

\section*{Acknowledgements}
ES and BDM are supported in part by the National Science Foundation (grant number AST-1615084) and the NASA Fermi Guest Investigator Program (grant number 80NSSC18K1708). ERC acknowledges support from NASA through the Einstein Fellowship Program, Grant PF6-170170. NCS and BDM are supported in part by the NASA Astrophysics Theory Program (grant number NNX17AK43G).

%%%%%%%%%%%%%%%%%%%%%%%%%%%%%%%%%%%%%%%%%%%%%%%%%%

%%%%%%%%%%%%%%%%%%%% REFERENCES %%%%%%%%%%%%%%%%%%

% The best way to enter references is to use BibTeX:

\bibliographystyle{mnras}
\bibliography{refs} % if your bibtex file is called example.bib

% Alternatively you could enter them by hand, like this:
% This method is tedious and prone to error if you have lots of references
%\begin{thebibliography}{99}
%\bibitem[\protect\citeauthoryear{Author}{2012}]{Author2012}
%Author A.~N., 2013, Journal of Improbable Astronomy, 1, 1
%\bibitem[\protect\citeauthoryear{Others}{2013}]{Others2013}
%Others S., 2012, Journal of Interesting Stuff, 17, 198
%\end{thebibliography}

%%%%%%%%%%%%%%%%%%%%%%%%%%%%%%%%%%%%%%%%%%%%%%%%%%

%%%%%%%%%%%%%%%%% APPENDICES %%%%%%%%%%%%%%%%%%%%%

%\appendix

%\section{Blah}

%%%%%%%%%%%%%%%%%%%%%%%%%%%%%%%%%%%%%%%%%%%%%%%%%%

% Don't change these lines
\bsp	% typesetting comment
\label{lastpage}
\end{document}